\documentclass[aps,prb,groupedaddress,showpacs,twocolumn,amsmath,amssymb]{revtex4}
\usepackage{graphicx}

\begin{document}

\title{Crossover between Hydrodynamic and Kinetic Modes in Binary Liquid Alloys}

\author{Stefano Cazzato$^1$, Tullio Scopigno$^1$, Taras Bryk$^2$, Ihor Mryglod$^2$ and Giancarlo Ruocco$^1$}
\email{tullio.scopigno@roma1.infn.it} \affiliation{$^1$INFM
CRS-SOFT, c/o Universit\'a di Roma ``La Sapienza'', ~I-00185,
Roma, Italy} \affiliation{$^2$Institute for Condensed Matter
Physics, National Academy of Sciences of Ukraine,
,UA-79011 Lviv, Ukraine}

%

\date{\today}

\begin{abstract}
Inelastic x-ray scattering (IXS) measurements of the dynamic
structure factor in liquid $\mathrm{Na_{57}K_{43}}$, sensitive to
the atomic-scale coarse graining, reveal a sound velocity value
exceeding the long wavelength, continuum value and indicate the
coexistence of two phonon-like modes. Applying Generalized
Collective Mode (GCM) analysis scheme, we show that the positive
dispersion of the sound velocity occurs in a wavelength region
below the crossover from hydrodynamic to atom-type excitations
and, therefore, it can not be explained as sound propagation over
the light specie (Na) network. The present result experimentally
proves the existence of positive dispersion in a binary mixture
due to a relaxation process, as opposed to fast sound phenomena.

\end{abstract}


\maketitle

\section{introduction}

Alkali metals in their liquid state are outstanding examples of
\emph{simple liquids}, as they encompass most of the physical
properties of complex fluids without system specific
complications. For this reason, in the last fifty years they have
been the subject of intensive investigations, both experimentally
and by means of computer simulations, aiming at an understanding
of the mechanisms underlying their collective dynamics at a
microscopic length scale \cite{coplov_rev,scop_rmp}. Interest in
the collective excitations and interdiffusion processes in liquid
binary mixtures, was stimulated by the pioneering Molecular
Dynamics (MD) study on the dynamics of the liquid Na-K alloy by
Jacucci and Mc Donald \cite{JAC_NaK}. Disparate masses metallic
alloys soon became the subject of intensive studies: the existence
of a new, high frequency mode with phase velocity exceeding the
hydrodynamic value was reported by MD simulations
\cite{JAC_LiPb1,JAC_LiPb,anento_lialloy}, and experimentally
proved by Inelastic Neutron Scattering (INS) in $\mathrm{Li_4Pb}$,
\cite{ver_LiPb_97,Ber_LiPb_98,ber_lipb}, in $\mathrm{NaCs}$
\cite{Gart_NaCs} and $\mathrm{Li_{30}Bi_{70}}$ \cite{bove_libi}.
The very nature of this additional mode, traditionally named
\textit{fast sound}, has been in focus of a lively debate in the
last three decades. It was either interpreted as an atom-type
acoustic mode, supported by the light (Li) ions only, which would
merge at low $Q$ with the corresponding low frequency mode (Pb)
into a single hydrodynamic velocity, or like an optic-like
excitation arising above a certain treshold $Q$ value. The
existence of a crossover from hydrodynamic into atom-type
excitations can be theoretically rationalized within a concept of
kinetic (non-hydrodynamic) modes in the framework of Generalized
Collective Modes (GCM) approach \cite{bryk_bin,bryk_hene}. It has
been shown that, at low $Q$, two modes exists and can be
associated to hydrodynamics density and concentration fluctuations
(collective region) while, above a certain characteristic
wavenumber (dependent on mass ratio), each of the two excitation
reflects the dynamics of a single atomic specie (atom-type
excitations). The experimental identification of the crossover,
however, has been heavily debated. This was mostly due to the
inherent difficulty of disentangling the coherent contribution
from the total INS scattering signal, paired by the kinematic
limitation restricting the accessible energy-momentum region. In
He-Ne mixtures, for instance, the upper limit of the hydrodynamic
region has been questioned, also in view of the possible
definitions of the excitation frequency in terms of the maxima of
the Dynamic Structure Factor (DSF), $S(Q,\omega)$, rather that
those of the longitudinal current autocorrelation function
\cite{baf_hene,pad_hene,baf_hene_r,sam_hene}.
\begin{equation}
C^L(Q,\omega)=\frac{\omega^2}{Q^2}S(Q,\omega)
\end{equation}
Similar controversies appeared in the case of water
\cite{tex_water,ber_water,sette_oldwater,sette_wat,sacchetti_wat},
where for long time the existence of fast sound was debated, until
the viscoelastic origin of such phenomenon was clarified
\cite{ruocco_water,monaco_water,fior_water}. In fact, the advent
of the new radiation sources opened the possibility to perform
Inelastic Scattering with X-rays (IXS), overcoming the previously
mentioned limitations of INS. The purely coherent X-ray atomic
cross section on one hand, and the relative high energy of the
probing X-rays on the other, allowed to study the purely coherent
dynamics of disordered systems over a wide energy-momentum range
covering the region across the first pseudo Brillouin region (i.e.
momentum transfers up to the first maximum of the static structure
factor, $Q_M$). As a consequence, the last decade saw a renewed
interest in the investigation of high frequency dynamics in
liquids and glasses.

It is worth to emphasize in this context the lack of a unique and
reliable methodology to analyze scattering experiments performed
on binary liquids. In majority of experimental studies one makes
use of either single DHO model \cite{demmelNaCl} or even a
memory-function ansatz -designed for one-component liquids-
 to estimate the dispersion law of collective
excitations in binary melts \cite{inui_cscl}. Within this kind of
approaches one necessarily ignores the existence of non-acoustic
high-frequency excitations in binary liquids, which beyond
hydrodynamic region contribute to the shape of partial dynamical
structure factors. In this study we show how one can apply more
sophisticated theoretical schemes to the analysis of experimental
data in a molten metallic alloy, namely a theoretical GCM
approach, which consistently treats contributions from
hydrodynamic and kinetic modes to dynamical processes in liquids.

We report here on an IXS study of the microscopic dynamics in the
molten alkali alloy $Na_{57}K_{43}$ which allowed us to:
\textit{i)} Identify the presence of two phonon-like modes, an
high frequency and a low frequency one. \textit{ii)} Point out, by
applying GCM theory to a binary mixture, the origin of these two
modes and how they contribute to the measured IXS spectra, i.e.
essentially to the mass density fluctuations spectra.
Specifically, we show the existence of a crossover from a
hydrodynamic regime, where only one of the two modes is active, to
an atom-type regime, where two excitations appear in the
$S(Q,\omega)$. \textit{iii)} Assign the sound velocity excess over
the hydrodynamic value to a relaxation process similar to that
observed in simple \cite{scop_rmp}, molecular \cite{mon_otpfq} and
hydrogen bonding liquids \cite{ruocco_water}, as opposed to the
idea of acoustic excitation propagating over the network supported
by the light specie or to the effect of optic-like excitations.

\section{the experiment}

The experiment reported in this work was carried out at the high
resolution beam line ID16 of the European Synchrotron Radiation
Facility (Grenoble, Fr). The backscattering monochromator and
analyzer crystals, operating at the $(11,11,11)$ silicon
reflections gave a total energy resolution of 1.5 meV, while
energy scans were performed by varying the temperature of the
monochromator with respect to that of the analyzer crystals. A
five analyzers bench allowed us to collect simultaneously spectra
at five different values of constant momentum transfer, covering a
$Q$ region below the position of the main diffraction peak
($Q_M\approx $18$ \mathrm{nm^{-1}}$). The sample consisted of an
NaK alloy at $T=300$ K with sodium concentration
$\mathrm{C_{Na}=57}$ atomic \%, and was hold in a silica capillary
of 2 mm inner diameter and 10 $\mathrm{\mu m}$ wall thickness,
sealed and kept in an inert atmosphere. Energy scans in the range
$-40<E<40$ $\mathrm{meV}$ where repeated up to a total integration
time of 300 s/point. Energy scans of the empty cell were collected
in the same $Q-E$ range as for the sample, prior to filling the
capillary, and were subtracted to the final IXS spectra.
\begin{figure}[h]
\centering
\includegraphics[width=.5\textwidth]{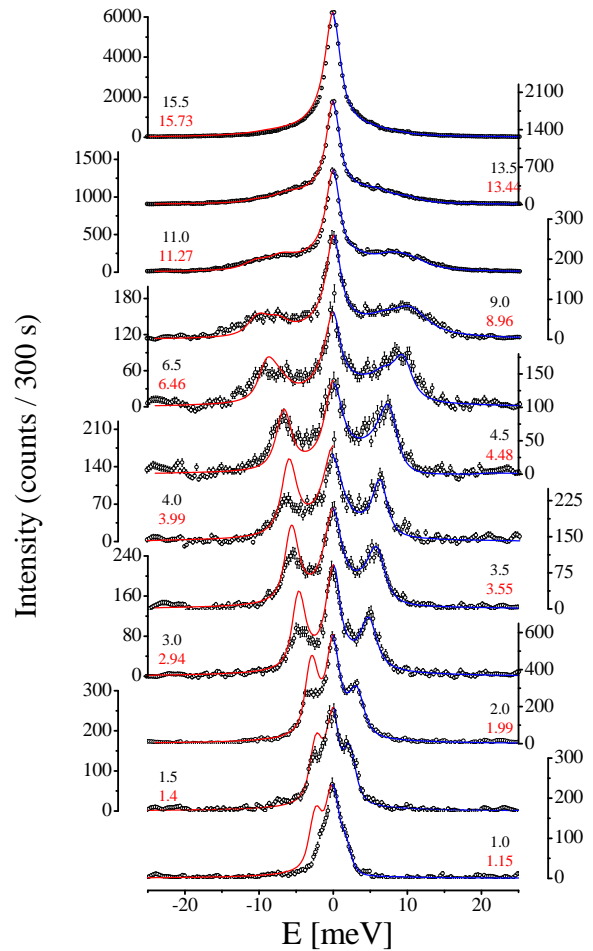}
 \caption{Selection of IXS spectra from the NaK liquid alloy
($c_{Na}=57 \mathrm{at\%}$, T=300K) measured for fixed $Q$ values
(nm$^{-1}$). The continuous line on the anti-Stokes side displays
the outcome of GCM theory, while on the Stokes side shows
Eq.\ref{eq:GCM} in which the eigenmodes are obtained as best fit
to the data, as explained in the text}\label{fig:panel}
\end{figure}
The signal measured in an IXS experiment is related to the double
differential cross-section depending on the exchanged momentum $Q$
and energy $E=\hbar \omega$, and it is basically the convolution
of the instrumental resolution, $R(\omega)$, with the classical
X-ray weighted DSF $S_{IXS}(Q,\omega)$,
\begin{equation}
I(Q,\omega)=\int{d\omega' R(\omega - \omega') S_{IXS}(Q,\omega')}
\frac{\hbar \omega' / KT }{1-e^{-\hbar \omega' / KT }}
\end{equation}
in which the last term accounts for the detailed balance
condition. The outcome of the IXS experiment (open circles) is
reported in Fig. \ref{fig:panel} for selected values of the
exchanged momentum $Q$.

In a binary mixture, $S_{IXS}(Q,\omega)$ can be represented as a
linear combination  of the partial dynamic structure factors
\begin{equation}
S_{ij}(Q,\omega)= \int dt e^{-i\omega
t}\left<n_i(Q,0)n_j^*(Q,t)\right>  (i,j=\mathrm{Na,K})
\end{equation}
with weights depending on the Na and K atomic form factors as well
as on the concentration \cite{bhatia1970}. In the previous
relation $n_i(Q,t)$ is the number density of species $i$.
Alternative representations of $S_{IXS}(Q,\omega)$ can be given in
terms of dynamical variables derived from the $n_i$ as independent
linear combinations. The most used options, in this respect, are
the Bathia Thornton \cite{bhatia1970} number-concentration, or the
March mass-concentration variable \cite{bhatia1974}. 
\section{theoretical model}

The usual approach to describe the DSF in a liquid is based on the
solution of the Langevin equation for an appropriate set of
dynamical variables. This can be done, for instance, by modelling
a suitable memory kernel. At least two relaxation times must be
accounted for in the second order memory function to describe the
density fluctuations in a monoatomic liquid \cite{scop_rmp}, and a
nontrivial generalization for the case of binary liquids is
required, along the line of the scheme proposed in Ref.
\cite{anento_lialloy}. For this reason we have chosen the GCM
approach, which additionally treats coupling with thermal
fluctuations and permits to solve the generalized Langevin
equation (GLE) in terms of dynamical eigenmodes \cite{bryk_hene}.

In such an approach the GLE is solved in a markovian
approximation, while correlations faster than the hydrodynamic
ones are accounted for by an appropriate choice for the set of
dynamical variables. On this ground the solution of the GLE is
straightforwardly given in terms of eigenvalues and eigenvectors
of the hydrodynamic matrix
\begin{equation}
\mathbf T(Q)=\mathbf F(Q,t=0)\cdot\left[\mathbf{\tilde
F}(Q,z=0)\right]^{-1}
\end{equation}
where $\mathbf F(Q,t)$ is the matrix of time correlation functions
of all the variables belonging to the set, and $\mathbf{\tilde
F}(Q,z)$ is its Laplace transform \cite{bryk_bin,bryk_hene}. The
partial DSFs $S_{ij}(Q,\omega)$ are then obtained by means of a
time Fourier transform of the corresponding components
$F_{ij}(Q,t)$ related, for example, to the partial number
densities $n_i$ and $n_j$. As we will show, the GCM approach
captures all the essential features of the reported IXS
experiment.

A suitable extended set of hydrodynamic variables for a binary
mixture is provided by the set \cite{bryk_bin}
\begin{equation}
A^{(8)}=\{n_t(Q,t),n_x(Q,t),J_t(Q,t),J_x(Q,t),\varepsilon(Q,t),\dot{J}_t(Q,t),\dot{J}_x(Q,t),\dot{\varepsilon}(Q,t)\}
\label{eq:8var}
\end{equation}
where the hydrodynamic variables reflecting the slowest
fluctuations in the binary liquid are: total number density
$n_t(Q,t)$, mass concentration density $n_x(Q,t)$, total
longitudinal mass current $J_t(Q,t)$ and energy density
$\varepsilon(Q,t)$. The variables with overdots correspond to the
first time derivatives of the relevant fluctuations and are
introduced to give a better representation of those process that
are faster than the hydrodynamic ones. It is worth to emphasize
here that, since total density and mass-concentration density can
be easily represented as linear combinations of partial densities,
one can represent the set (\ref{eq:8var}) via partial dynamical
variables. Hence, the generalized hydrodynamic matrix $\mathbf
T(Q)$ constructed on t-x dynamical variables (\ref{eq:8var}) or on
partial dynamical variables will have identical eigenvalues,
because both sets of dynamical variables are connected by linear
transformation.

Within this scheme, the IXS weighted DSF can be expressed as:
\begin{equation}
S_{IXS}(Q,\omega)=\frac{1}{\pi}Re \left [
 \sum_{1}^{8} \frac{G_{\alpha}^{IXS}(Q)}{z+z_{\alpha}(Q)} \right]_{z=i\omega}
\label{eq:GCM}
\end{equation}
 Both the eigenvalues ($z_{\alpha}(Q)$, which are complex or
real in the case of propagating or diffusive modes, respectively)
and the corresponding eigenvectors ($G_{\alpha}^{IXS}(Q)$ which
are the X-ray weighted eigenvectors, i.e. are determined as the
appropriate linear combination of partial Na and K densities,
accounting for form factors and atomic concentration), can be
evaluated within GCM theory once the $Q$-dependent elements of the
hydrodynamic matrix are known. This requires the knowledge of
initial values of time correlation functions of the variables set,
as well as the correlation times of the kind
\begin{equation}
\tau_{mn}(Q)=\frac{1}{F_{mn}(Q,0)}\int_0^{\infty}F_{mn}(Q,t)dt
\label{eq:tempi}
\end{equation}
with $~m,n=n_t,n_x,\varepsilon $ \cite{bryk_hene}. To evaluate
these latter we performed molecular dynamics simulations on a
system of 8000 particles interacting via effective two-body
potentials \cite{wax_nak} in a microcanonical ensemble at
$T=298$K. The production run was performed over $3\times 10^5$
timesteps, and all the static and time correlation functions
needed for GCM analysis were directly calculated in MD. The
minimal wavenumber reached was $0.81$ $nm^{-1}$. The resulting
spectrum consists of eight dynamical eigenmodes:
complex-conjugated pairs of eigenvalues correspond to phonon-like
collective excitations $z_\alpha(Q)=\sigma_\alpha(Q)\pm
i\omega_\alpha(Q)$ with $\omega_\alpha(Q)$ and $\sigma_\alpha(Q)$
being frequency and damping of $\alpha$-th excitation,
respectively, while real eigenvalues represent purely diffusive
relaxation processes.

The prediction of parameter-free GCM theory (red lines) are
reported in Fig. \ref{fig:panel} along with the lineshape of
Eq.\ref{eq:GCM} in which the eigenmodes are obtained through a
best fit to the experimental data (blue lines). In this latter
case we used as initial values those from GCM, and kept fixed all
the frequencies of the propagating eigenvalues (imaginary part of
the complex eigenvalues), and the (real) eigenvalues of the
diffusive modes (diffusion coefficient). The weight coefficients
$G^{IXS}_{\alpha}(Q)$ as well as the damping coefficients (real
parts) of propagating complex eigenvalues were instead let as free
parameters. Sum rules where used as constraints to further reduce
the number of free parameters \footnote{ we used the first three
sum rules: $\sum_{\alpha}G_{\alpha}(Q)=S_{IXS}(Q)$;
$\sum_{\alpha}G_{\alpha}(Q)z_{\alpha}(Q)=0$;
$\sum_{\alpha}G_{\alpha}(Q)z_{\alpha}^2(Q)=-\omega_0^2(Q)$ }. The
stage of fitting procedure is necessary since the GCM analysis is
based on MD simulations, which use effective two-body potentials.
Namely the use of effective potentials can introduce some extra
error, because it is well known, that effective two-body
potentials very rarely can yield the correct melting point,
therefore under- or overestimating the contributions from
relaxation processes to the shape of dynamical structure factors.
In the anti-Stokes side of Fig. \ref{fig:panel} one can see, that
the parameter-free GCM approach correctly predicts the frequency
of collective excitations observed in the scattering experiments,
only relative weight of relaxation processes (central peak) and
propagating modes is not well reproduced. Namely at that point the
aforementioned fitting procedure was applied, and later we will
show, that the fitting procedure does not alter the main results
of GCM analysis.
\begin{figure}[h]
\centering
\includegraphics[width=0.5\textwidth]{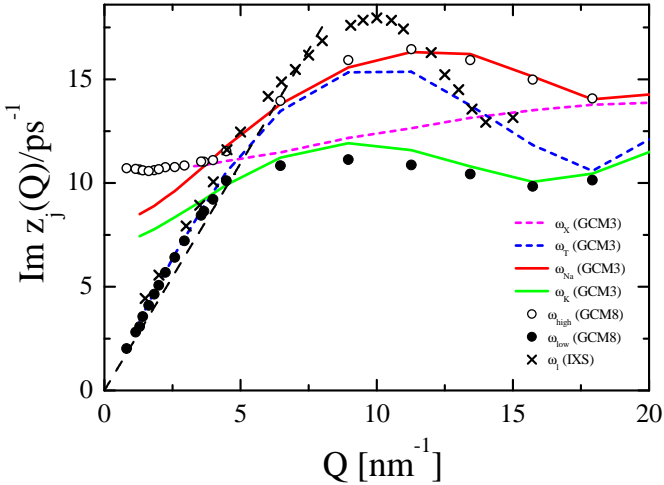}
\vspace{-6.4 cm}\caption{Experimental dispersion relation obtained
from the maxima of $C_{IXS}(Q,\omega)$ ($\times$), along with the
hydrodynamic dispersion derived by ultrasound measurements
\cite{amaral1974} ($---$). The two modes predicted by complete
8-variables GCM theory (high frequency $\circ$, low frequency
$\bullet$), and the outcome of partial 3-variables GCM analysis
with total (blue - - -) concentration (pink - - -), Na (red
------) and K (green ------) variables are also
reported.}\label{fig:dispersion}
\end{figure}
Figure \ref{fig:dispersion} shows the imaginary part of the two
complex eigenvalues of the 8-variables GCM treatment (circles and
dots). In order to ascertain the nature of these phonon-like
modes, we performed additional GCM analysis using four
projected-out onto processes of different origin 3-variable
eigenvalue problems using basis sets
\begin{eqnarray}
A^{(3t)}&=&\{n_t(Q,t),J_t(Q,t),\dot{J}_t(Q,t)\} \nonumber \\
A^{(3x)}&=&\{n_x(Q,t),J_x(Q,t),\dot{J}_x(Q,t)\} \nonumber \\
A^{(3Na)}&=&\{n_{Na}(Q,t),J_{Na}(Q,t),\dot{J}_{Na}(Q,t)\} \nonumber \\
A^{(3K)}&=&\{n_{K}(Q,t),J_{K}(Q,t),\dot{J}_{K}(Q,t)\}
\label{eq:GCM3}
\end{eqnarray}
This approach allows to ascertain the origin of each branch of
collective excitations in the spectrum and processes responsible
for them in different Q-regions
\cite{bryk_opt,bryk_bin,bryk_nacl}. The result is reported in the
same Fig. \ref{fig:dispersion} and clearly shows how the atom type
excitations (dotted blue and pink lines for the Na and K subsets,
respectively) correspond to the two eigenvalues of the 8-variable
treatment in the high Q limit, while at low Q the same eigenvalues
are reproduced by the total density and concentration subsets.
This is a clear indication of the existence of two dynamical
regimes, a low $Q$, collective region, and an high $Q$, atom-type
region, in agreement with recent observation of optic-like modes
above $\approx 2$ nm$^{-1}$ in MD simulations on Li alloys
\cite{anento_lialloy}. Interestingly, the experimental sound
velocity $c_l(Q)=\omega_l(Q)/Q$ determined by the maximum
$\omega_l$ of the raw current autocorrelation function $\omega^2
S_{IXS}(Q,\omega)$ (black crosses) turns out to be mainly
determined by the low energy mode at low $Q$ and by the high
energy mode at high $Q$. In addition to that, it does not reach,
even at the lowest explored $Q$, the hydrodynamic (adiabatic)
value \cite{amaral1974}, while exceeds it systematically
\footnote{alternative definitions of the sound velocity have been
recently discussed \cite{baf_hyd}, in this context we just stress
that both definitions lead to the same conclusion, i.e.
higher-than hydrodynamic values}.
\begin{figure}[h]
\centering \vspace{-.8 cm}
\includegraphics[width=0.43\textwidth]{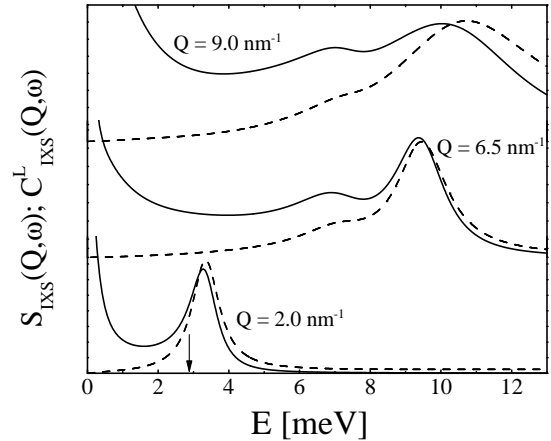}
\vspace{-4 cm}\caption{Classical resolution deconvoluted
$S(Q,\omega)$ (continuous line) and $C^L(Q,\omega)$ (dashed lines)
obtained from the IXS measurement. The presence of a single mode
exceeding the adiabatic frequency (marked by the arrow) is clearly
visible at Q=2 nm$^{-1}$, i.e. well below the crossover to
atom-type dynamics. For Q=6.5 nm$^{-1}$ two excitations appears,
with the high frequency one dominating at Q=9.0
nm$^{-1}$.}\label{fig:dec}
\end{figure}
This is clearly shown in Fig. \ref{fig:dec} in which we report the
DSF, $S(Q,\omega)$ and longitudinal current spectra,
$C^L(Q,\omega)$, from Eq.(\ref{eq:GCM}) adjusting the weight of
the high frequency and low frequency modes to the IXS spectra. At
the lowest Q values a single mode dominates the spectrum, with an
energy clearly exceeding the one expected from hydrodynamic
dispersion $\omega=c_0 Q$. Around the first crossover (Q=6.5
$nm^{-1}$) both the high frequency and the low frequency mode do
contribute to the spectra, while at $Q=9$ $nm^{-1}$ most of the
IXS signal is due to the high frequency, Na-like excitation.

\begin{figure}[h]
\centering
\includegraphics[width=.4\textwidth]{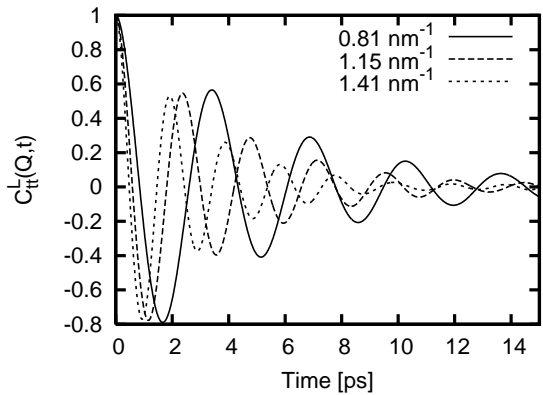}
 \caption{Normalized total mass current autocorrelation function at the three lowest Q values investigated in MD}\label{fig:cltt}
\end{figure}

\begin{figure}[h]
\centering
\includegraphics[width=.4\textwidth]{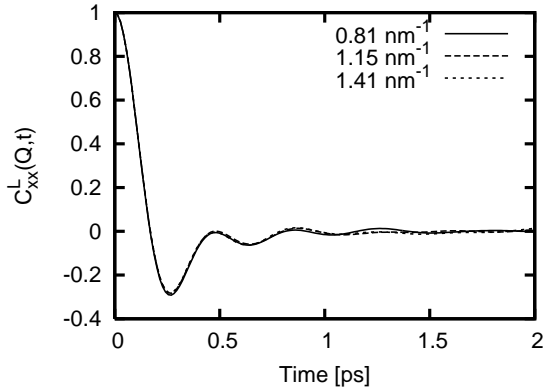}
 \caption{Normalized mass-concentration current autocorrelation function at the three lowest Q values investigated in MD}\label{fig:clxx}
\end{figure}

\begin{figure}[h]
\centering
\includegraphics[width=0.45\textwidth]{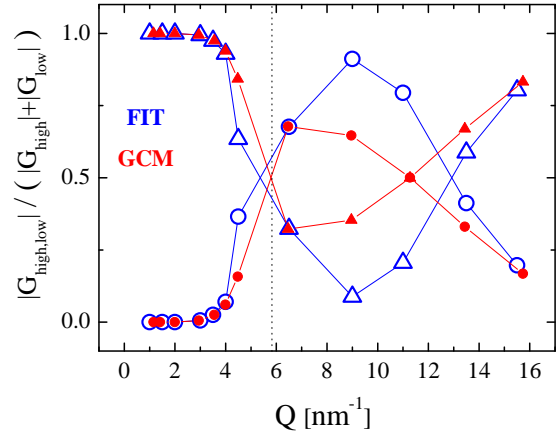}
\vspace{-5.1 cm}\caption{High (circle) and low (triangle)
frequency modes from GCM (red) and from fitting procedure (blu).
The crossover at $Q\approx 6$ nm$^{-1}$ can be
observed.}\label{fig:weights}
\end{figure}

This picture is substantiated by direct inspection of the MD data.
Different time correlation functions derived in MD simulations and
analyzed by the GCM approach can reveal different types of
collective excitations existing in binary liquids. In Figs.
\ref{fig:cltt} and \ref{fig:clxx} we show MD-derived longitudinal
total and mass-concentration autocorrelation function at three
lowest wavenumbers sampled in simulations. It is clearly seen,
that the mass-concentration current autocorrelation functions
$C^L_{xx}(Q,t)$ functions reflect propagating excitations with
much smaller time scale, than the sound excitations, which as in
the case of single-component liquids, contribute to the shape of
$C^L_{tt}(Q,t)$ functions. Furthermore, almost identical shape of
$C^L_{xx}(Q,t)$ functions at small wavenumbers implies, that the
short-time excitations have almost the same frequency and damping
in the long-wavelength limit, that is usually the evidence of
kinetic optic-like excitations
The presence of the two phonon-like modes in the IXS spectra, at
intermediate $Q$ values, can be conveniently quantified looking at
the relative weights reported in Fig.\ref{fig:weights}.
\begin{figure}[h]
\centering
\includegraphics[width=.44\textwidth]{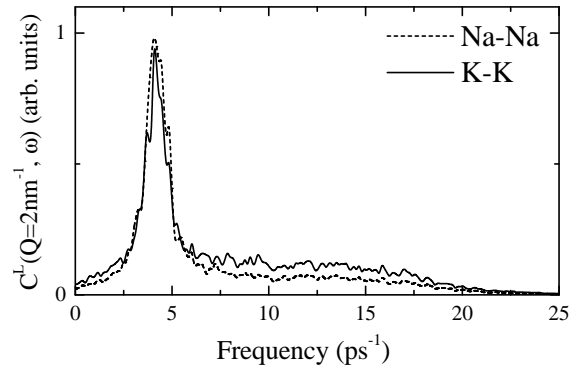}
\vspace{-5.6 cm} \caption{MD calculated $C^L_{NaNa}(Q,\omega)$ (-
- -) and $C^L_{KK}(Q,\omega)$ (-----) for Q=2nm$^{-1}$. The same
acoustic mode turns out to dominate in both species, and no sign
of fast sound-like phenomenology, in which the light component is
dominated by the high frequency mode, is recovered at Q lower than
the crossover value ($Q\sim 5 nm^{-1}$, see Fig.
\ref{fig:mddispersion}) from MD.} \label{fig:mdspectra}
\end{figure}
The low frequency mode is clearly dominant below the sharp
crossover occurring at $Q\approx 6$ nm$^{-1}$. Around this value
both modes contribute to the IXS spectra, while above the high
frequency mode dominates up to a new crossover at $Q\approx 13$
nm$^{-1}$. The IXS cross section is indeed roughly proportional to
the total density autocorrelation function, and hence below the
crossover it reflects the collective longitudinal excitation and
not the optic-like mode. As previously mentioned, the amplitude
values from GCM and from fitting procedure are quite different,
but the oscillating behavior with the presence of sharp crossovers
is captured by both the methods. The fact that positive dispersion
observed from the IXS data occurs well below the first crossover,
when the spectral frequency is solely reproduced by the low
frequency mode, allows us to rule out a ``fast sound" mechanism
i.e. the transition from the hydrodynamic regime to a dynamical
regime in which the sound propagates over a network formed by the
light component of the mixture, as it happens in disparate masses
alloys. On the contrary, the origin of such an effect has to be
traced back to a relaxation process of viscoelastic origin,
similar to that invoked in simple \cite{scop_rmp} and molecular
\cite{mon_otpfq} liquids, glasses \cite{gcr_prlsim,scop_presim}
and in the controversial case of water and hydrogen bonding system
\cite{ruocco_water,monaco_water,ange_HF,fior_water}.
\begin{figure}[h]
\centering
\includegraphics[width=.47\textwidth]{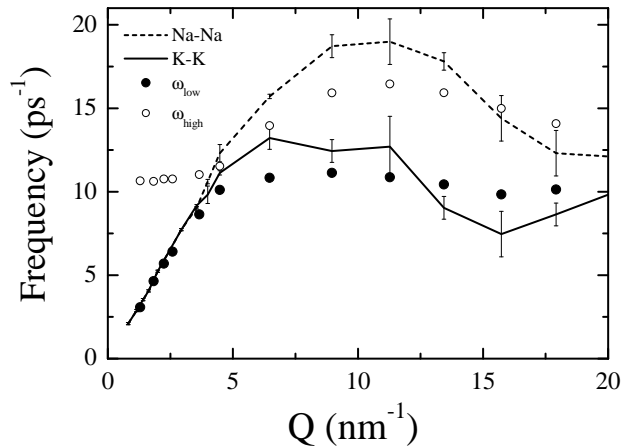}
\vspace{-5.5 cm} \caption{MD calculated dispersion relation for Na
(- - -) and K (-----), the high ($\circ$) and low ($\bullet$)
frequency GCM modes are also reported. At Q's below the crossover
the same scenario depicted in Fig. \ref{fig:mdspectra} holds for
the partial $C^L_{NaNa}(Q,\omega)$ and $C^L_{KK}(Q,\omega)$
currents, which are dominated by what can be addressed to as the
GCM low frequency mode.
 }\label{fig:mddispersion}
\end{figure}
Along the same line, the inspection of the MD derived spectral
functions $C^L_{NaNa}(Q,\omega)$ and $C^L_{KK}(Q,\omega)$
presented in Fig. \ref{fig:mdspectra} for a $Q=2$ nm$^{-1}$- i.e.
well below the crossover- is particularly enlightening. The
fingerprint of fast sound would be here the presence of an extra
high frequency mode in the light component (Na) spectra which, in
turn, is absent in the corresponding heavy component (K) spectra.
As can be observed in Fig. \ref{fig:mdspectra}, this is certainly
not the present case. Conversely, the same acoustic mode dominates
in both the species, accompanied by a common weak ($Q^2$
vanishing) high frequency feature. As shown in Fig.
\ref{fig:mddispersion}, this scenario holds at any $Q$'s below the
crossover, while at high Q's the partial spectra are dominated by
distinct optic-like modes which can be identified with the high
and the low frequency GCM modes.

\section{conclusions}
In conclusion, we reported the results of an experimental IXS
investigation of high frequency dynamics in the liquid binary
mixture $\mathrm{Na_{57}K_{43}}$ which suggest the presence of two
distinct phonon-like excitations beside diffusive modes. A
quantitative analysis based on GCM theory allows to identify the
nature of these excitations. The relative weight of the
phonon-like modes shows an oscillatory behavior with $Q$, ruled by
sharp crossovers defining collective and atom-type dynamic
regions, in which the global density/concentration fluctuation and
the partial Na/K density fluctuations dominate, respectively. The
observed positive dispersion of the acoustic branch, i.e. a sound
velocity value exceeding the long wavelength limit, occurs well
within the collective region. This indicates for this process an
origin due to the existence of a microscopic relaxation,
ubiquitous in monoatmic liquids, as opposed to the fast sound
phenomena observed in disparate masses mixtures, where enhanced
sound velocity has been suggested to be either due to the
propagation of sound over the light component network or to the
presence of an optic branch.

\section{acknowledgements}
We acknowledge valuable assistance during the experiment by R. Di
Leonardo and the IXS team at the ESRF.


\end{document}